\documentstyle[12pt]{article}

\advance\textwidth 1cm
\advance\oddsidemargin -0.5cm
\def\ft#1#2{{\textstyle{{#1}\over{#2}}}}
\def\vdm2{\Delta^2}
\def\de{\delta}
\def\res{\mathop{\rm Res}}

\newcommand\beq{\begin{equation}}
\newcommand\eeq{\end{equation}}
\def\beqa{\begin{eqnarray}}
\def\eeqa{\end{eqnarray}}
\def\bega{\begin{array}}
\def\enda{\end{array}}
\def\non{{\nonumber}}

\def\bl{{\biggl(}}
\def\br{{\biggr)}}

\def\m{{\mu}}
\def\d{\partial}

\def\a{\alpha}
\def\b{\beta}

\def\ga{\gamma}
\def\g{\gamma}

\begin{document}
\title{An elegant solution of\\the n-body Toda problem}

\author{Arlen Anderson\thanks{arley@physics.unc.edu}\\
Isaac Newton Institute\\
20 Clarkson Road\\
Cambridge CB3 0EH, England\\
and\\
Blackett Laboratory\\
Imperial College\\
Prince Consort Rd.\\
London SW7 2BZ, England\\
and\\
Dept. Physics\thanks{Permanent address}\\
Univ. North Carolina\\
Chapel Hill NC 27599-3255 USA}
\date{July 18, 1995}
\maketitle
\vspace{-16cm}
\hfill UNC-IFP-95-513

\hfill hep-th/9507092
\vspace{16cm}

\begin{abstract}
The solution of the classical open-chain n-body Toda problem is derived
from an ansatz and is found to have a highly symmetric form.  The proof
requires an unusual identity involving Vandermonde determinants.  The
explicit transformation to action-angle variables is exhibited.
\end{abstract}

\newpage
\baselineskip 24pt

The Toda chain is one of the paradigmatic examples of an
integrable many-body system of interacting particles.  The discovery of
its conserved integrals of motion\cite{Tod,Fla} and its subsequent
solution\cite{Kac,Mos,Min} were important steps in the development of the
theory of integrable systems\cite{Tod2}.  An almost universal feature of
analytical studies of the Toda system is the use of the Lax pair
formalism.  In this paper, an alternative derivation of the solution of
the classical open-chain n-body Toda system is given.

The derivation proceeds essentially from an ansatz about the form of the
solution and therefore lacks the power and generality of the Lax pair
treatment. The solution however has an elegant structure which is not
evident in previous representations. More, it can be interpreted as the
classical canonical transformation from the Toda system to a free theory.
This is an important clue to constructing the classical and quantum
solutions by a sequence of elementary canonical
transformations\cite{And1}. Following the successful solution of the
three-body Toda problem with this approach\cite{ANPS}, work is in progress
on the classical and quantum open-chain n-body problems.

The Hamiltonian for the (n+1)-body open chain Toda system is
\beq
H=\ft12 \sum_{k=1}^{n+1} p_k^2 + \sum_{k=1}^n e^{q_k-q_{k+1}}.
\eeq
The arguments of the exponential potentials can be interpreted as
expressions for the root vectors of $A_n$ in the Cartan basis\cite{Min}. A
coordinate transformation will put the root vectors into the Chevalley
basis and separate out the motion of the center of mass. The transformation
is given by
\beqa
\label{com}
q_1&\mapsto& q_1 +{q_{n+1}\over n+1}, \\
q_k&\mapsto& -q_{k-1}+q_k+{q_{n+1}\over n+1},\quad\quad(2\le k\le n) \non \\
q_{n+1}&\mapsto& -q_n+{q_{n+1}\over n+1}, \non \\
p_k&\mapsto& {1\over n+1}\bl \sum_{j=k}^{n} (n+1-j) p_j -
\sum_{j=1}^{k-1}  j p_j \br+ p_{n+1},\quad\quad (1\le k\le n). \non\\
p_{n+1}&\mapsto& {-1\over n+1}\bl
\sum_{j=1}^{k-1}  j p_j \br+ p_{n+1}. \non
\eeqa
The transformed Hamiltonian is
\beqa
\label{ham}
H^a&=&{1\over 2(n+1)} \bl \sum_{k=1}^n k(n+1-k) p_k^2 + \sum_{k=2}^n
\sum_{j=1}^{k-1} 2j(n+1-k) p_j p_k \br +{n+1\over 2} p_{n+1}^2 \non \\
&&\quad\quad +e^{2q_1-q_2} +\sum_{k=2}^{n-1} e^{2q_k -q_{k-1} -q_{k+1}} +
e^{2q_n-q_{n-1}}.
\eeqa
This leads to the equations of motion
\beqa
\label{eom}
\ddot q_1&=& -e^{2q_1-q_2} \\
\ddot q_k&=& -e^{2q_k-q_{k-1}-q_{k+1}} \quad\quad (2\le k\le n-1) \non \\
\ddot q_n&=& -e^{2q_n-q_{n-1}}. \non
\eeqa
The solution of these equations has the remarkably simple form
\beq
\label{sol}
e^{-q_m}=\sum_{j_1<\cdots<j_m}^{n+1} f_{j_1}\cdots f_{j_m}
\vdm2(j_1,\ldots,j_m) e^{(\m_{j_1}+ \cdots+ \m_{j_m})t},
\eeq
where $\vdm2(j_1,\ldots,j_m)$ is the square of the Vandermonde determinant
\beq
\label{vdm}
\vdm2(j_1,\ldots,j_m)=\prod_{j_i<j_k} (\m_{j_i}-\m_{j_k})^2,
\eeq
and $f_{k}$ and $\m_k$ are arbitrary constants, satisfying
\beqa
\label{constr}
\prod_{k=1}^{n+1} f_k &=& \Delta^{-2}(1,\dots,n+1), \\
\sum_{k=1}^{n+1} \m_k &=& 0. \non
\eeqa
(There are additional constraints on the range of the $f_k$ if one
requires the $q_m$ be real.)
The solution has $2n$ free parameters as required.  The solution in
the original variables is determined from the transformation (\ref{com})
to be composed of ratios of these solutions times a factor for
the center of mass motion.

To derive the solution, make the ansatz
\beq
e^{-q_m}=\sum_{j_1<\cdots<j_m}^{n+1} f_{j_1\cdots j_m} e^{(\m_{j_1}+
\cdots +\m_{j_m})t},
\eeq
where the $\m_k$ are arbitrary real numbers.  Note that this ansatz
defines a variable
\beq
e^{-q_{n+1}}=f_{1\cdots n+1}e^{(\m_1+\cdots+\m_{n+1}) t}.
\eeq
Such a variable might naturally appear in the final equation of (\ref{eom})
to give $\ddot q_n= -e^{2q_n-q_{n-1}-q_{n+1}}$. That this variable does not
appear can be interpreted as meaning that one has set $q_{n+1}=0$.
This ultimately is the origin of the restrictions (\ref{constr}) on
the $f_k$ and
$\m_k$.  The equation for $\ddot q_1$ is also of the form of the others
if there is a $q_0=0$.  The open-chain Toda system thus has fixed
endpoints in this sense.
The ansatz and solution are compatible with the slightly more general problem
where $e^{-q_{n+1}}=ce^{\kappa t}$.  Then, in the solution, one would have
$\prod_{k=1}^{n+1} f_k=c\Delta^{-2}(1,\dots,n+1)$ and
$\sum_{k=1}^{n+1} \m_k=\kappa$.

Consider $e^{-q_m}$.
Differentiating twice and multiplying by $e^{-q_m}$ leads to
\beq
-\ddot q_m e^{-2q_m}= e^{-q_m} \d_t^2 e^{-q_m}  - (\d_t e^{-q_m})^2.
\eeq
But from the equations of motion $-\ddot q_m
e^{-2q_m}=e^{-q_{m-1}-q_{m+1}}$ (using $q_0=0=q_{n+1}$).
Substituting the ansatz into the resulting equation gives ($2\le m\le n-1$)
\beqa
\label{eq1}
\sum_{{j_1<\cdots <j_m \atop {k_1<\cdots <k_m \atop j_1<k_1}}}^{n+1}
f_{j_1\cdots j_m}f_{k_1\cdots k_m}
(\sum_{i=1}^m \m_{j_i}-\sum_{i=1}^m \m_{k_i})^2
e^{(\sum_{i=1}^m \m_{j_i} +\sum_{i=1}^m \m_{k_i})t} &=& \\
&&\hspace{-8cm}= \sum_{{j_1<\cdots <j_{m-1} \atop k_1<\cdots <k_{m+1}}}^{n+1}
f_{j_1\cdots j_{m-1}}f_{k_1\cdots k_{m+1}} e^{(\sum_{i=1}^{m-1} \m_{j_i} +
\sum_{i=1}^{m+1} \m_{k_i})t}.\non
\eeqa
The equation for $m=1$ is
\beq
\label{eq2}
\sum_{j_1<j_2}^{n+1} f_{j_1} f_{j_2} (\m_{j_1}-\m_{j_2})^2
e^{(\m_{j_1}+\m_{j_2})t} =\sum_{j_1<j_2}^{n+1} f_{j_1 j_2}
e^{(\m_{j_1}+\m_{j_2})t}.
\eeq
The equation for $m=n$ involves $f_{j_1\cdots j_n}$ where $1\le j_1<
\cdots <j_n\le n+1$.  As one is choosing $n$ integers out of $n+1$,
this is more succinctly labelled by $f_{\hat r}$ where $r$ is the
integer which is not in the set. Similarly $f_{\widehat{rs}}$ means the
two integers $r\ne s$ do not appear, and the indices of $f$ are the
remaining $n-1$ integers.  With this notation, the equation for $m=n$ is
\beq
\label{eq3}
\sum_{r<s}^{n+1} f_{\hat r}f_{\hat s}(-\m_{r}+\m_{s})^2 e^{(-\m_r-\m_s
+2\sum_{k=1}^{n+1} \m_k)t} =
\sum_{r<s}^{n+1} f_{\widehat{rs}} e^{(-\m_r-\m_s+\sum_{k=1}^{n+1}\m_k)t}.
\eeq

Assume the $\m_k$ are all distinct and that they have no accidental
degeneracies in their linear combinations,
such as $\m_{j_1}+\m_{j_2}=\m_{j_3}+\m_{j_4}$. The asymptotic behavior
of the exponentials can be used to equate like terms in the sums.
The degenerate cases can be recovered later by continuity in the $\m_k$.
Let
\beq
\label{fs}
f_{j_1\cdots j_m}=f_{j_1}\cdots f_{j_m}\vdm2(j_1,\ldots,j_m),
\eeq
where the $f_{j_k}$ are (so far) arbitrary constants and
$\vdm2(j_1,\ldots,j_m)$ is the square of the
Vandermonde determinant (\ref{vdm}).
With this definition, the $m=1$ equation (\ref{eq2}) is easily verified.
The $m=n$ equation (\ref{eq3}) is satisfied if
the constraints (\ref{constr}) on the $f_k$ and $\m_k$ are imposed.
The proof that Eq. (\ref{eq1}) is satisfied reduces to
a heirarchy of identities for Vandermonde determinants.

On the left hand side of (\ref{eq1}), there are two sets of
indices $\{j_\a\}$ and $\{k_\b\}$.  Since $j_1<k_1$, at most they can
have $m-1$ indices in common.   The asymptotic behavior of the
exponential is given by a sum over the $\m_i$ indexed by the
combined set $S=\{j_\a,k_\b \}$.  Different partitions of $S$ into
sets $\{j_\a\}$ and $\{k_\b\}$ ($j_1<k_1$) will have the
same asymptotic behavior.  The number of such terms will depend on
the number of distinct indices between the two sets, and these
constitute separate cases.  Let $2r$ denote the number of distinct
indices.

Consider the case $r=1$, labelling the common indices $s_1,\dots, s_{m-1}$
and the distinct ones $j_1$ and $k$ ($j_1<k$).  There is a unique term on
both sides of (\ref{eq1}) with the asymptotic behavior given by this
set of indices, and one has
\beq
f_{j_1s_1\cdots s_{m-1}} f_{\{k s_1\cdots s_{m-1}\}} (\m_{j_1}-\m_k)^2
=f_{s_1\cdots s_{m-1}} f_{\{j_1ks_1\cdots s_{m-1}\}},
\eeq
where the curly brackets indicate the indices should be arranged in
increasing order.  Using (\ref{fs}), the constant factors $f_i$ cancel
and one has a relation between Vandermonde determinants
\beqa
\vdm2(j_1,s_1,\ldots,s_{m-1}) \vdm2(k,s_1,\ldots,s_{m-1})(\m_{j_1}-\m_k)^2
&=&\\
&&\hspace{-3.5cm}=\vdm2(s_1,\ldots,s_{m-1}) \vdm2(j_1,k,s_1,\ldots,s_{m-1}).
\non
\eeqa
Using the relation
\beq
\vdm2(k,s_1,\ldots,s_{m-1})=\vdm2(s_1,\ldots,s_{m-1}) \prod_{i=1}^{m-1}
(\m_k-\m_{s_i})^2
\eeq
and its relatives, the dependence on the common indices is seen to cancel
and one is left with the identity
\beq
\label{r1}
(\m_{j_1}-\m_k)^2=\vdm2(j_1,k).
\eeq

It is a general feature  for all $r$ that the dependence on the constant
factors and the common indices cancels on both sides, so without loss of
generality one can focus on the distinct indices alone.  Reindex the
set $S$ of distinct indices by the integers 1 to $2r$.  Partition
$S$ into two sets $\a=\{1,\a_2,\ldots,\a_r\}$ and $\b=\{\b_1,\ldots,\b_r\}$
and denote the collection of such partitions $P_{\a\b}$.  Separately
partition $S$ into sets $\ga=\{\g_1,\ldots,\g_{r-1}\}$ and
$\de=\{\de_1,\ldots,\de_{r+1}\}$, calling the collection of partitions
$P_{\ga,\de}$.  Denote $\vdm2(\a;r)=\vdm2(1,\a_2,\ldots,\a_r)$ and
similarly for the rest.  The number $r$ of indices involved in
the Vandermonde determinant is made explicit to reduce confusion.
Both sides of Eq. (\ref{eq1}) will be equal if the following
identity between Vandermonde determinants holds
\beq
\label{id}
\sum_{P_{\a\b}} \vdm2(\a;r) \vdm2(\b;r) (\sum_\a \m_\a -
\sum_\b \m_\b)^2 = \sum_{P_{\ga \de}} \vdm2(\ga;r-1) \vdm2(\de;r+1).
\eeq

It seems likely that this identity has a group theoretical interpretation,
but in its absence, the identity can be proved inductively as
follows\cite{God}.  Divide both sides by $\vdm2(S;2r)$.  This gives
the equation
\beq
\sum_{P_{\a\b}} { (\sum_\a \m_\a - \sum_\b \m_\b)^2  \over
\prod_{\a,\b} (\m_\a-\m_\b)^2} = \sum_{P_{\ga \de}}
{1\over \prod_{\ga,\de} (\m_\ga-\m_\de)^2}.
\eeq
Denote the left hand side of the equation
by $L_r$ and the right hand side by $R_r$.  The equation $L_1=R_1$ holds
trivially. Assume that $L_{r-1}=R_{r-1}$.
The inductive step will be made by considering the pole structure of
$L_r$ and $R_r$.  Since $L_r$ and $R_r$ are analytic functions of the
$\m_i$ without zeroes, if they can be shown to have the same residue at
all of their poles, they must be equal.

Choose two indices from the set $S$, neither equal to 1,
and let their associated $\m_i$ be labelled $z$ and $a$.  (The index 1
is special because it has a preferred role in the partitioning.
Which of the original $\m_i$ is associated to the index 1 is however
arbitrary, so one can investigate the pole structure at the $\m_i$ missed
here by reindexing the set $S$.) Let $S'$ denote the set $S$ with these two
indices removed, and let $\a',\b',\ga',\de'$ denote partitions of $S'$
as defined above with $r$ replaced by $r-1$.

Consider the residue of $L_r$ at $z=a$.  $L_r$ has
a double pole at $z=a$ if $z\in\a$ and $a\in\b$ or {\it vice versa}.
In the former case, the residue is computed to be
\beqa
\res_{z=a} L_r\biggr|_{z\in\a,a\in\b} &=&\\
&&\hspace{-3cm}= {2 \over
\prod_{S'} (a-\m_{S'})^2}
\sum_{P_{\a' \b'}}\bl {1\over \sum_{\a'} \m_{\a'} - \sum_{\b'} \m_{\b'} }
- \sum_{\b'} {1\over a-\m_{\b'}} \br {(\sum_{\a'} \m_{\a'} -
\sum_{\b'} \m_{\b'})^2 \over \prod_{\a',\b'} (\m_{\a'}-\m_{\b'})^2}.\non
\eeqa
In the alternative case $z\in\b$, $a\in\a$, the residue is
\beqa
\res_{z=a} L_r\biggr|_{z\in\b,a\in\a} &=& \\
&&\hspace{-3.2cm}={2\over
\prod_{S'} (a-\m_{S'})^2}
\sum_{P_{\a' \b'}}\bl -{1\over \sum_{\a'} \m_{\a'} - \sum_{\b'} \m_{\b'} }
- \sum_{\a'} {1\over a-\m_{\a'}} \br{(\sum_{\a'} \m_{\a'} -
\sum_{\b'} \m_{\b'})^2 \over \prod_{\a',\b'} (\m_{\a'}-\m_{\b'})^2}. \non
\eeqa
Adding these, the residue of $L_r$ at $z=a$ is
\beq
\res_{z=a} L_r=-{2L_{r-1} \over
\prod_{S'} (a-\m_{S'})^2} \sum_{S'} {1\over a-\m_{S'}}.
\eeq

The residue at $z=a$ of $R_r$ is similarly composed of terms where
$z\in\ga$,  $a\in\de$ and {\it vice versa}.  The full residue is
\beq
\res_{z=a} R_r=-{2R_{r-1} \over
\prod_{S'} (a-\m_{S'})^2} \sum_{S'} {1\over a-\m_{S'}}.
\eeq
This is seen to equal the residue of $L_r$ at $z=a$, given $L_{r-1}=R_{r-1}$.
Since this result holds for all pairs of the original $\m_i$, one concludes
that $L_r=R_r$ and the induction is complete.

To exhibit the solution (\ref{sol}) as a canonical transformation from
(\ref{ham}) to a Hamiltonian independent of coordinates, one must introduce
final coordinates and momenta and find a
relation between them and the $f_j$ and $\m_k$ so that the transformation is
canonical.  It is clear that one can redefine $f_j$ by an overall constant,
\beq
f_j=e^{\bar x_j} \tilde f_j.
\eeq
The arguments of the exponentials then define the final coordinates
\beq
x_j=\m_j t+\bar x_j.
\eeq
There should only be $n$ independent degrees of freedom, and
the coordinate $x_{n+1}=-\sum_{i=1}^n x_i$ is not independent because
it is related
to the others by the constraints (\ref{constr}).
It is useful however to introduce a temporary form of the final Hamiltonian
\beq
\tilde H={1\over 2}\sum_{i=1}^{n+1} \m_i^2.
\eeq
The $\m_j$ are not the momenta conjugate to $x_j$ because if the
the constraint $\m_{n+1}=-\sum_{i=1}^n \m_i$ were eliminated, the
wrong $\dot x_j$ would follow from Hamilton's equations.  It is necessary
to introduce $n$ momenta $k_j$ conjugate to the $x_j$, so that $\dot x_j=
{\d \tilde H\over \d k_j}=\m_j$.  The relation
between $k_j$ and $\m_j$ is found to be
\beq
k_j=\m_j -\m_{n+1}=\m_j +\sum_{i=1}^n \m_i
\eeq
or in reverse ($j\ne n+1$)
\beqa
\label{mk}
\m_j &=& k_j-{1\over n+1} \sum_{i=1}^n k_i, \\
\m_{n+1} &=& -{1\over n+1} \sum_{i=1}^n k_i \non
\eeqa
The final Hamiltonian is then
\beq
\label{ham2}
\tilde H={n\over 2(n+1)}\sum_{j=1}^n k_j^2 -{1\over n+1}\sum_{i<j} k_i k_j.
\eeq

The next step is to find an equation for the evolution of the original
momenta.  This is easily done by taking a time derivative of the solution
(\ref{sol})
\beq
\label{sol1}
e^{-q_m}=\sum_{j_1<\cdots<j_m}^{n+1} \tilde f_{j_1}\cdots \tilde f_{j_m}
\vdm2(j_1,\ldots,j_m) e^{x_{j_1}+ \cdots+ x_{j_m}}
\eeq
to find
\beq
-\dot q_m e^{-q_m}=
\sum_{j_1<\cdots<j_m}^{n+1} \tilde f_{j_1}\cdots \tilde f_{j_m}
\vdm2(j_1,\ldots,j_m) (\m_{j_1}+\cdots +\m_{j_m})e^{x_{j_1}+ \cdots+ x_{j_m}}.
\eeq
Using Hamilton's equations with the Hamiltonian (\ref{ham}), one can
express $\dot q_m$ in terms of the momenta as
\beq
\dot q_m={1\over (n+1)}[m(n+1-m) p_m + \sum_{i=1}^{m-1} i(n+1-m) p_i
+\sum_{i=m+1}^n m(n+1-i)p_i],
\eeq
The result is
\beqa
\label{sol2}
{-1\over (n+1)}[m(n+1-m) p_m + \sum_{i=1}^{m-1} i(n+1-m) p_i
+\sum_{i=m+1}^n m(n+1-i)p_i]e^{-q_m} &=& \non \\
&&\hspace{-12cm} =\sum_{j_1<\cdots<j_m}^{n+1} \tilde f_{j_1}\cdots
\tilde f_{j_m}
\vdm2(j_1,\ldots,j_m) (\m_{j_1}+\cdots+ \m_{j_m})e^{x_{j_1}+ \cdots+ x_{j_m}}.
\eeqa

Finally, by requiring that the Poisson brackets be preserved under the
transformation, one can determine the $\tilde f_j$ in terms of the $k_i$'s.
The result is that ($j\ne n+1$)
\beqa
\label{fk}
\tilde f_j &=& (-1)^{j-1} k_j^{-1} \prod_{i\ne j}^n (k_j-k_i)^{-1}, \\
\tilde f_{n+1} &=& \prod_{i=1}^n k_i^{-1} \non
\eeqa
One confirms that the $f_j$ satisfy the constraint (\ref{constr}). (Note
that the maximal symmetry is evident in terms of the $\m_i$'s since
$k_j=\m_j-\m_{n+1}$ and $k_j-k_i=\m_j-\m_i$.)  The proof that this is the
correct form for the $f_j$ follows by constructing the Poisson brackets
and collecting like exponentials.  Conditions are quickly found that the $f_j$
must be particular products of differences between momenta.  It is then
seen that there are no additional requirements.

Using (\ref{mk}) and (\ref{fk}) in (\ref{sol1}) and (\ref{sol2}) gives the
explicit canonical transformation between the open-chain n-body Toda
Hamiltonian in the
Chevalley basis (\ref{ham}) and a Hamiltonian (\ref{ham2}) which is
independent of coordinates.  The reduction to action-angle variables is
essentially complete.  From this point, one can attempt to construct a
product of elementary canonical transformations which produces this full
transformation.  This has been done for the 3-body system \cite{ANPS} and
work is in progress on the n-body system.  The value of such a product
is that, when it is found in the quantum system, it
allows the construction of integral representations of the eigenfunctions
of the system.

Acknowledgements.  I would like to thank P. Goddard for suggesting the
proof of the identity relating Vandermonde determinants.  This work was
supported in part by National Science Foundation grant PHY 94-13207.

\end{document}